\documentclass[12pt]{article}
\usepackage{epsf}
\usepackage{amsmath}
\usepackage{graphics}
\usepackage{cite}

\renewcommand{\thefootnote}{\#\arabic{footnote}}
\begin{document}

\newcommand{\gtrsim}{ \mathop{}_{\textstyle \sim}^{\textstyle >} }
\newcommand{\lesssim}{ \mathop{}_{\textstyle \sim}^{\textstyle <} }

\newcommand{\rem}[1]{{\bf #1}}

\renewcommand{\thefootnote}{\fnsymbol{footnote}}
\setcounter{footnote}{0}
\begin{titlepage}

\def\thefootnote{\fnsymbol{footnote}}

\begin{center}
\hfill astro-ph/0607391\\
\hfill September 2006\\
\vskip .5in
\bigskip
\bigskip
{\Large \bf Dark Matter Candidate from Conformality}

\vskip .45in

{\bf P.H. Frampton}

{\em Institute of Field Physics, Department of Physics and Astronomy,}

{\em University of North Carolina, Chapel Hill, NC 27599-3255, USA}

\end{center}

\vskip .4in
\begin{abstract}

Abelian quiver gauge theories provide candidates for
the conformality approach to physics beyond the standard model which
possess novel cancellation mechanisms for quadratic divergences. 
A $Z_2$ symmetry (R parity) can be imposed
and leads naturally to a dark matter candidate which is the Lightest
Conformality Particle (LCP), a neutral spin-1/2 state
with weak interaction annihilation cross section, mass in the 100 GeV 
region and relic density of non-baryonic
dark matter $\Omega_{dm}$ which can be consistent with the observed
value $\Omega_{dm} \simeq 0.24$.
\end{abstract}
\end{titlepage}

\renewcommand{\thepage}{\arabic{page}}
\setcounter{page}{1}
\renewcommand{\thefootnote}{\#\arabic{footnote}}

\newpage

\bigskip

\noindent {\it Introduction}

\bigskip

One approach to the hierarchy, or naturalness, problem is to postulate
conformality, four-dimensional conformal invariance at high energy, for
the non gravitational extension of the standard model.
The conformality
approach suggested\cite{PHF1998} in 1998 has made considerable progress. 
Models which contain the standard model fields have been constructed\cite{Z7}
and a model which grand unifies at about 4 TeV\cite{Z12} has been
examined.

Such models are inspired by the AdS/CFT correspondence\cite{Maldacena,Witten}
specifically based on compactification of the IIB superstring
on the abelian orbifold $AdS_5 \times S^5/Z_n$ with N coalescing
parallel D3 branes. A model is specified by N and by the embedding
$Z_n \subset SU(4)$ which is characterized by integers
$A_m$ ($m = 1,2,3,4$) which specify how the {\bf 4} of $SU(4)$
transforms under $Z_n$.
Only three of the $A_m$ are independent because
of the $SU(4)$ requirement that $\Sigma_m A_m = 0$ (mod n).
The number of vanishing $A_m$ is the
number ${\cal N}$ of surviving supersymmetries. Here we
focus on the ${\cal N}=0$
case.

In \cite{CFR}, the original speculation \cite{PHF1998}
that such models may be conformal has been refined to
exclude models which contain scalar fields
transforming as adjoint representations because only if all
scalars are in bifundamentals are there chiral fermions and,
also only if all scalars are in bifundamentals, the
one-loop quadratic divergences cancel in the scalar propagator.
I regard it as encouraging that these two desirable
properties select the same subset of models.

Another phenomenological encouragement stems from the
observation\cite{Vafa} that the standard model representations for
the chiral fermions can all be accommodated in bifundamentals
of $SU(3)^3$ and can appear naturally in the conformality
approach. 

In the present article I address the issue of dark matter. From
studies of galactic rotation curves, large scale structure,
cosmic microwave background, high redshift supernovae and
other observational cosmology there is strong evidence
of the need for non-baryonic dark matter. One of the
most recent estimates is from WMAP3\cite{WMAP3} which
finds a break down of the overall energy density of
$4\%$ baryons, $72\%$ dark energy and $24\%$ non-baryonic dark matter.

Here I define a $Z_2$ symmetry for the conformality 
theory in the following subsection. I then show that it leads to an
Lightest Conformality Particle (LCP)
which is an attractive candidate for the dark
matter particle.

\newpage

\bigskip

\noindent {\it Definition of a $Z_2$ symmetry} 

\bigskip

In the quiver gauge theories, the
gauge group, for abelian orbifold $AdS_5 \times S^5/Z_n$
is $U(N)^n$. In phenomenological application $N=3$
and $n$ reduces eventually after symmetry breaking
to $n=3$ as in trinification.
The chiral fermions are then in the representation of $SU(3)^3$:
\begin{equation}
(3, 3^*, 1) + (3^*, 1, 3) + (1, 3, 3^*)
\end{equation}
This is as in the {\bf 27} of $E_6$ where the particles
break down in to the following representations of the
$SU(3) \times SU(2) \times U(1)$ standard model group:
\begin{equation}
Q, ~~~~~ u^c, ~~~~ d^c, ~~~~ L ~~~~ e^c ~~~~~ N^c
\end{equation}
transforming as
\begin{equation}
(3,2), ~~ (3^*,1), ~~ (3^*,1), ~~ (1,2), ~~ (1,1), ~~ (1,1) \\
\end{equation}
in a {\bf 16} of the $SO(10)$ subgroup. In addition there are the states
\begin{equation}
h, ~~~~~ h*, ~~~~~ E, ~~~~~ E*
\end{equation}
transforming as
\begin{equation}
(3, 1), ~~ (3^*, 1), ~~ (2,1), ~~ (2,1)
\end{equation}
in a {\bf 10} of $SO(10)$
and finally
\begin{equation}
S
\end{equation}
transforming as the singlet
\begin{equation}
(1, 1)
\end{equation}

It is natural to define a $Z_2$ symmetry $R$ which commutes
with the $SO(10)$ subgroup of $E_6 \rightarrow O(10) \times U(1)$
such that $R=+1$ for the first {\bf 16} of states. 
Then it is mandated that $R=-1$ for the {\bf 10} and {\bf 1}
of SO(10) because the following Yukawa couplings must
be present to generate mass for the fermions:
\begin{equation}
16_f 16_f 10_s, ~~~~16_f 10_f 16_s, ~~~~ 10_f 10_f 1_s, ~~~~
10_f 1_f 1_s, ~~~~ 1_f 1_f 1_s
\end{equation}
which require the $R$ assignments for the scalars $R=+1$
for $10_s, 1_s$ and $R=-1$ for $16_s$. In E(6)
various possibilities for R parity including this one
were analysed in \cite{Ma}.

\bigskip

The LCP is the lightest linear combination of the three neutral
components of E, E* and S. It is expected to have mass $\sim 1$ TeV
and is a WIMP candidate for dark matter.

\newpage

\bigskip

\noindent {\it Annihilation Cross-Section}

\bigskip

The LCP act as cold dark matter WIMPs, and the calculation of the
resultant energy density follows a well-known path. Here
I follow the procedure in \cite{GKJ}.

The LCP decouple at temperature $T_*$, considerably less than their
mass $M_{LCP}$; I define $x_* = M_{LCP}/T_*$. Let the
annihilation cross-section of the LCP
at decoupling be $\sigma_*$. Then the dark matter density
$\Omega_{dm}$, relative to the critical density,
is estimated as
\begin{equation}
\Omega_{dm} h_{75}^2 = \frac{\tilde{g}_*^{1/2}}{g_*} x_*^{3/2} \left(
\frac{3 \times 10^{-38} cm^2}{\sigma_*} \right)
\label{OmegaM}
\end{equation}
where $h_{75}$ is the Hubble constant in units
of $75km/s/Mpc$. 
$g_* = (g_b + \frac{7}{8}g_f)$ is the effective number of 
degrees of freedom (d.o.f.) at freeze-out for all particles
which later convert their energy into photons;
and $\tilde{g}_*$ is the number
of d.o.f. which are relativistic at $T_*$. 

\bigskip
\bigskip

Thus, to estimate the non-baryonic dark matter density
arising from LCPs, I need estimates
of five quantities occurring in Eq.(\ref{OmegaM}):
$h_{75}, \tilde{g}_*, g_*, x_*$ and $\sigma_*$, and
to this I now turn.

\bigskip
\bigskip

I start with $h_{75}$ where the central value from WMAP3\cite{WMAP3}
is $H_0 = 72 km/s/Mpc$ and so a 
good estimate of $h_{75}$ is $h_{75} = 72/75 = 0.96$.

\bigskip
\bigskip

For the energy ranges I consider, $\tilde{g}_*=g_*$ and depends
on the freeze-out temperature $T_*$. We consider masses in the range
$30 GeV \leq M_{LCP} \leq 2 TeV$. Since $x_* = M_{LCP}/T_*$ is
relatively insensitive to $M_{LCP}$, as we shall see shortly,
always within the values $20 \leq x_* \leq 30$,
the freeze-out temperatures of relevance will be in the range
$1 GeV \leq T_* \leq 100 GeV$.

For these $T_*$ we compute:

\bigskip

For $100 GeV \geq T_* \geq 10 GeV$:
\begin{equation}
g_* = 86.25 
\end{equation}

\bigskip

For $10 GeV \geq T_* \geq 3 GeV$:
\begin{equation}
g_* = 75.75 
\end{equation}

\bigskip

For $3 GeV \geq T_* \geq 1 GeV$:
\begin{equation}
g_* = 61.75 
\end{equation}

\bigskip

\newpage

\bigskip
\bigskip

The value of $x_*$ may be estimated using the formula\cite{Kolb}

\begin{eqnarray}
X & = & 0.038 \sqrt{g_*} M_{Planck} M_{LCP} \sigma_* \\
x_* & = & ln X - \frac{1}{2}ln ln X
\label{xstar}
\end{eqnarray}
I have already estimated $g_*$. I use $M_{Planck} = 10^{19} GeV$.

\bigskip
\bigskip

The annihilation cross section $\sigma_*$ for LCPs at freeze-out
may be estimated using analogs of the Feynman graphs used 
in \cite{Griest}. A naive estimate of $\sigma_*$
follows from the dimensional formula\cite{Gunn}
$\sigma_* \sim G_F^2 T_*^2$, but we shall use a more
detailed calcualtion, see Eq.(\ref{sigmaZ}) below.
From Eq.(\ref{OmegaM}) and the estimates for $h_{75},
g_*, \tilde{g}_*, x_*$ already given, the cross-section
must satisfy $\sigma_* \geq 3 \times 10^{-35} cm^2$
with the lower bound saturating $\Omega_{dm}$ and
a smaller cross-section being
unacceptable because it leads to too much dark matter.
Empirical bounds \cite{ALEPH} require $M_{LCP} \geq 43.1 GeV$. 
The allowed range is generically \cite{ALEPH,DreesNojiri,Franke,Pandita,Pandita2}
$43.1 GeV \leq M_{LCP} \leq 1 TeV$ (see below).

\bigskip

One important contributing Feynman graph is $Z$ exchange in the direct
channel which gives, by itself, the cross-section
\begin{equation}
\sigma_* (X\bar{X} \rightarrow f\bar{f})
= \frac{1}{128\pi  M_X^2} [\alpha^2 + \beta^2]^2\left[ 
\frac{g_2(M_Z)^4}{16 \cos^4\theta_W(M_Z)} \frac{M_f^2 M_X^2}{M_Z^4} \right]
\label{sigmaZ}
\end{equation}
in which $\alpha, \beta$ are defined by
\begin{equation}
\Phi_{LCP} = \alpha E^0 + \beta \bar{E}^0 + \gamma S^0
\label{LCP}
\end{equation}
so that $\alpha, \beta$ are coefficients of doublets and $\gamma$ is coefficient of a singlet.

\bigskip

Let the mass of the fermion $f$ be $M_f = f_{10} \times 10 GeV$. Then using $\alpha_2(M_Z) 
=g_2(M_Z)^2/4\pi = 0.0339$, $sin^2 \theta_W(M_Z) = 0.231$, $M_Z=91.19 GeV$,
$1 (GeV)^{-2} \equiv 3.894 \times 10^{-28} cm^2$,  leads,
independently of $M_X = M_{LCP}$, to $\sigma_*(X\bar{X} \rightarrow f\bar{f})
 = 2.68 \times 10^{-38} [\alpha^2 + \beta^2]^2 \times (f_{10})^2 ~~ cm^2$.

\bigskip
 
For the top quark $f \equiv t$ we find for $f_{10} = 17.2$
that $\sigma_* = 6.61 \times 10^{-36} [\alpha^2 + \beta^2]^2 cm^2$.
To generate all the dark matter we require
\begin{equation}
f_{10} [\alpha^2 + \beta^2] \leq 36.1
\end{equation}
which suggests for reasonable values (not very close to pure singlet) $M_{LCP} < 1 TeV$.

\bigskip
\bigskip

\newpage

\noindent {\it Fermion mass hierarchy}

\bigskip
\bigskip

In the conformality approach, the Yukawa couplings at the conformal scale, usually
$\sim 4$ TeV, are of order one. Thus, when the electroweak $SU(2) \times U(1)$ is
broken it is natural that all quarks and charged leptons would acquire mass
comparable to the wek scale. Although this is valid for the top quark, all
the other fermion masses in the standard model are smaller; this is the
fermion hierarchy problem which also effects estimation of the LCP mass.

\bigskip

Conformality does not predict this fermion hierarchy but can {\it accommodate} it
by adding soft mass terms after breaking $U(3)^n \rightarrow U(3)^3 \rightarrow
SU(3) \times SU(2) \times U(1) \rightarrow SU(3)\times U(1)$ (For details of the gauge symmetry breaking
see the trinifiaction analysis in \cite{Glashow,Pakvasa,Willenbrock}
\footnote{Note that the trinification here has the major difference from
that proposed in \cite{Glashow} and studied in
\cite{Pakvasa,Willenbrock} that unification is at a Teravolt,
not a Yottavolt, scale.}). The soft terms must be 
fine-tuned significantly to cancel the mass acquired in gauge symmetry breaking.
For the up and down quarks such tuning is $1$ part in $10^4$ while for the electron
it is $1$ part in $250,000$. Regretfully alternative approaches 
have no more predictivity about masses than here.

\bigskip

Because of this conceptual question, it is
merely assumed that the LCP is at $\sim 100$ GeV; there is every reason to believe this
is a possible outcome. An improved understanding of {\it spontaneous} breaking
of conformal symmetry using the ideas of \cite{PHFMPLA} may shed light on the mass spectrum.

\bigskip

\noindent {\it Discussion}

\bigskip

The LCP is a viable candidate for a
cold dark matter particle which can be produced at the LHC.
To produce all of the nonbaryonic cold dark matter
$\Omega_{LCP}=\Omega_{dm} \simeq 0.24$ requires that
the mass of the LCP be in the range
$43.1 GeV(ALEPH) \leq M_{LCP} \leq 1 TeV$.

\bigskip

The distinction from other dark matter candidates
will require establishment of the $U(3)^3$ gauge bosons, extending the 3-2-1 standard model
and the discovery that the LCP is in a bifundamental
representation thereof.

\bigskip

To confirm that the LCP is the dark matter particle would, however,
require direct detection of dark matter. In bolometric experiments
with a small number events the LCP will appear similar
to other WIMPs but with high statistics the unique couplings
of the LCP will be distinguishable. 

\bigskip

It has been established that conformality
can provide (i) naturalness without one-loop
quadratic divergence for the scalar mass \cite{CFR} and
anomaly cancellation\cite{DF};
(ii) precise unification of the coupling constants \cite{Z12,FRT};
and (iii) a viable dark matter candidate. 
It remains for experiment to determine whether
quiver gauge theories with gauge group $U(3)^3$ or $U(3)^n$ with $n \geq 4$
are employed by Nature.

\vspace{1.0cm}

\begin{center}

{\bf Acknowledgements}

\end{center}

\bigskip
I thank my colleagues E. Di Napoli and R. Rohm for discussions.
This work was supported in part by the
U.S. Department of Energy under Grant No. DE-FG02-97ER-41036.

\end{document}